\documentclass[pra,aps,amsmath,amssymb,amsfonts,twocolumn,nofootinbib,longbibliography,superscriptaddress]{revtex4-2}
\usepackage{amssymb}
\usepackage{bm,mathrsfs}
\usepackage{graphicx}
\usepackage{amsmath,bbm}
\usepackage{amsfonts,amssymb}
\usepackage{times}
\usepackage{verbatim}
\usepackage{amsmath}
\usepackage{bm}
\usepackage{float}
\usepackage[colorlinks,breaklinks,linkcolor=blue,anchorcolor=blue,citecolor=blue,urlcolor=blue]{hyperref}
\allowdisplaybreaks
\begin{document}
 \title{Connection-topology--dependent energy transport in quantum battery networks with reciprocal and nonreciprocal couplings}
	\author{Bing-Bing Liu}
    	\affiliation{Quantum Information Institute, School of Physics and Laboratory of Zhongyuan Light,
Zhengzhou University, Zhengzhou 450001, China}
     \author{Rui Chen}
	\affiliation{School of Physics and Technology, Central China Normal University, Wuhan, Hubei 430072, China}
    \author{Jin-Lei Wu}
	\affiliation{Quantum Information Institute, School of Physics and Laboratory of Zhongyuan Light,
Zhengzhou University, Zhengzhou 450001, China}
        \author{Gang Chen}\email{chengang971@163.com}
	\affiliation{Quantum Information Institute, School of Physics and Laboratory of Zhongyuan Light,
Zhengzhou University, Zhengzhou 450001, China}
 \author{Shi-Lei Su}\email{slsu@zzu.edu.cn}	\affiliation{Quantum Information Institute, School of Physics and Laboratory of Zhongyuan Light,
Zhengzhou University, Zhengzhou 450001, China}
\affiliation{Institute of Quantum Materials and Physics, Henan Academy of Science, Henan 450046, China}
\date{\today}
        
\begin{abstract}

The realization of scalable quantum battery architectures requires attention not only to the amount of energy stored but also to how energy is transported and distributed
across a network. While previous studies focused on collective charging in multi-cell quantum batteries, topology-dependent transport laws in quantum battery networks remain less explored. Here we investigate quantum battery networks with
reciprocal and engineered nonreciprocal couplings and compare cascaded and parallel architectures within a unified transport framework. In the nonreciprocal regime, the optimal coupling follows distinct topology-dependent scaling laws, $J_{\rm op}^{c}\propto N$ for cascaded transport and $J_{\rm op}^{p}\propto N^{-1/2}$ for parallel charging in the large-$N$ limit. In reciprocal cascaded networks, a parity-dependent spectral response produces an odd-even transport effect that is absent in the nonreciprocal and parallel configurations. We further analyze the role of thermal and squeezed reservoirs and show that thermal noise mainly increases passive energy, whereas squeezing enhances the useful fraction of the stored energy, thereby increasing the ergotropy. These results shift the emphasis from charging enhancement to transport engineering and provide architecture-level design principles for quantum battery networks.
	\end{abstract}

\maketitle
\section{Introduction} 
Quantum batteries, i.e., quantum systems engineered to store and release energy, have attracted considerable attention owing to the potential advantages offered by quantum coherence, correlations, and many-body effects in enhancing charging speed, charging power, and energy-storage capacity~\cite{PhysRevE.87.042123,Binder_2015,RevModPhys.96.031001,Ferraro2026,PhysRevLett.129.130602,PhysRevLett.124.130601,PhysRevB.98.205423,PhysRevLett.125.180603,Hymas2026,rkyk-14dj,10.1116/5.0184903}. Beyond optimizing the stored energy, ergotropy is also a key metric, which quantifies the maximum work extractable through unitary operations~\cite{Allahverdyan_2004,PhysRevLett.122.047702,PhysRevLett.131.030402}. A practically useful quantum battery should be assessed not only by how much energy is accumulated, but also by how efficiently that energy can be delivered and converted into extractable work.

A variety of charging protocols have been proposed, including feedback control charging~\cite{Mitchison2021,PhysRevApplied.19.064069,PhysRevResearch.7.013151}, stimulated Raman adiabatic
passage~\cite{PhysRevE.100.032107,Dou2021,Hu_2022}, optimal quantum control~\cite{Rodrguez_2024,10.1063/5.0161354}, among others~\cite{doi:10.1126/sciadv.abk3160,PhysRevLett.131.240401,PhysRevB.108.L180301,PhysRevA.107.032203,PhysRevA.111.042216,PhysRevA.106.032212,PhysRevA.109.052206}. Multi-cell quantum batteries have been widely investigated for collective effects leading to superextensive charging advantages~\cite{PhysRevLett.118.150601,PhysRevLett.128.140501,PhysRevLett.133.197001,PhysRevA.110.032205,PhysRevB.99.205437,PhysRevA.97.022106}, including Dicke and Sachdev–Ye–Kitaev models, with two-photon Dicke batteries providing further power enhancement~\cite{PhysRevLett.120.117702,PhysRevE.99.052106,PhysRevB.105.115405,PhysRevA.109.022210,PhysRevLett.125.236402,PhysRevResearch.6.023136,PhysRevB.102.245407,PhysRevB.109.235432}. However, quantum batteries are inevitably affected by decoherence and exhibit self-discharge effects~\cite{PhysRevA.100.043833}. Such detrimental influences can be mitigated by dark states~\cite{PhysRevApplied.14.024092} and Floquet engineering~\cite{PhysRevA.102.060201}. In particular, Ref.~\cite{d9k1-75d4} shows that the unique hyperfine interaction of the $^{14}\rm N$ nucleus can enhance the ratio of coherent ergotropy, thereby improving the robustness against self-discharge.

In contrast to focusing on suppressing dissipation, noise-assisted and environment
engineered mechanisms have emerged as effective routes for enhancing charging performance and mitigating degradation~\cite{PhysRevLett.122.210601,PhysRevLett.132.090401,Huang2021,Li2024,PhysRevE.104.064143,PhysRevResearch.2.033413,Kamin_2020,PhysRevA.102.052223,PhysRevA.109.022226,PhysRevE.105.064119,PhysRevA.104.032207,PhysRevA.103.033715,PhysRevE.104.044116,vqnk-kzqg}. Nonreciprocity generated by cooperative dissipation has emerged as a particularly promising mechanism because it enables
broadband directional transport and suppresses backflow~\cite{PhysRevX.5.021025,PRXQuantum.4.010306,doi:10.1126/sciadv.adj8796,PhysRevLett.132.120401,PhysRevLett.126.223603}. This mechanism has already been exploited to enhance
the performance of quantum batteries~\cite{PhysRevLett.132.210402,43n6-rnj3,kh36-7z76,PhysRevApplied.23.024010,fbv7-m7sd,Lin_2026}. Nonreciprocity has also been introduced into the Dicke model and applied to enhance multi-cell batteries~\cite{PhysRevLett.131.113602,fn1b-2m9g}.

However, previous studies primarily focused on quantum batteries composed of multiple cells; the role of network connectivity among batteries has not been fully explored. In quantum technologies, energy storage is not expected to rely on a single
battery unit; instead, multiple batteries are generally expected to be interconnected to form quantum battery networks~\cite{LUAN2021102896,8610191}. Such architectures are closer to practical quantum energy systems and naturally give rise to energy transport processes between different batteries, leading to richer dynamical behavior beyond that of isolated battery cells. Meanwhile, related works have explored energy transport in cascaded and parallel spin ensembles~\cite{PhysRevB.104.L140303,jy9l-l8hv}, as well as ergotropy transport in a one-dimensional spin chain~\cite{96cm-ktcb}.

The present work addresses this next layer of the problem. Rather than focusing only on charging enhancement, we investigate how topology, transport directionality, and reservoir properties jointly determine energy delivery, steady-state energy storage, and ergotropy in quantum battery networks. We consider both cascaded and parallel architectures and derive analytical results that reveal their distinct transport characteristics. In particular, cascaded networks exhibit sequential delivery, while parallel networks exhibit collective redistribution. This distinction leads to opposite optimal-coupling trends with the battery number $N$: $J_{\rm op}^{c}\propto N$ and $J_{\rm op}^{p}\propto N^{-1/2}$ in the large-$N$ limit. We emphasize that our analysis establishes a general framework: the regimes discussed in Refs.~\cite {fbv7-m7sd,PhysRevLett.132.210402} correspond to the weak- and strong-coupling parameter regimes, respectively. More recently, the ultrastrong-coupling regime has also been explored~\cite{liu2026dissipative}.

We further show that the reciprocal cascaded network exhibits a parity-dependent spectral transport mechanism, which explains the odd-even dependence of the terminal-battery response and is absent in the nonreciprocal and parallel cases. Finally, reservoir properties are also important in determining the performance of quantum batteries~\cite{PhysRevA.108.052213,b5lx-j66h,Li_2026}. By analyzing both thermal and squeezed reservoirs, we show that stored energy and ergotropy exhibit distinct responses: thermal noise primarily increases the passive energy, whereas squeezing enhances the useful fraction of the stored energy. The present results provide insight into the mechanisms governing energy transport in quantum battery networks and may offer useful guidance for the design of scalable quantum energy storage architectures and future quantum energy networks.

\section{Transport model and optimal operating point}
\begin{figure}
		\centering
  \includegraphics[width=8.5cm,height=6.2cm]{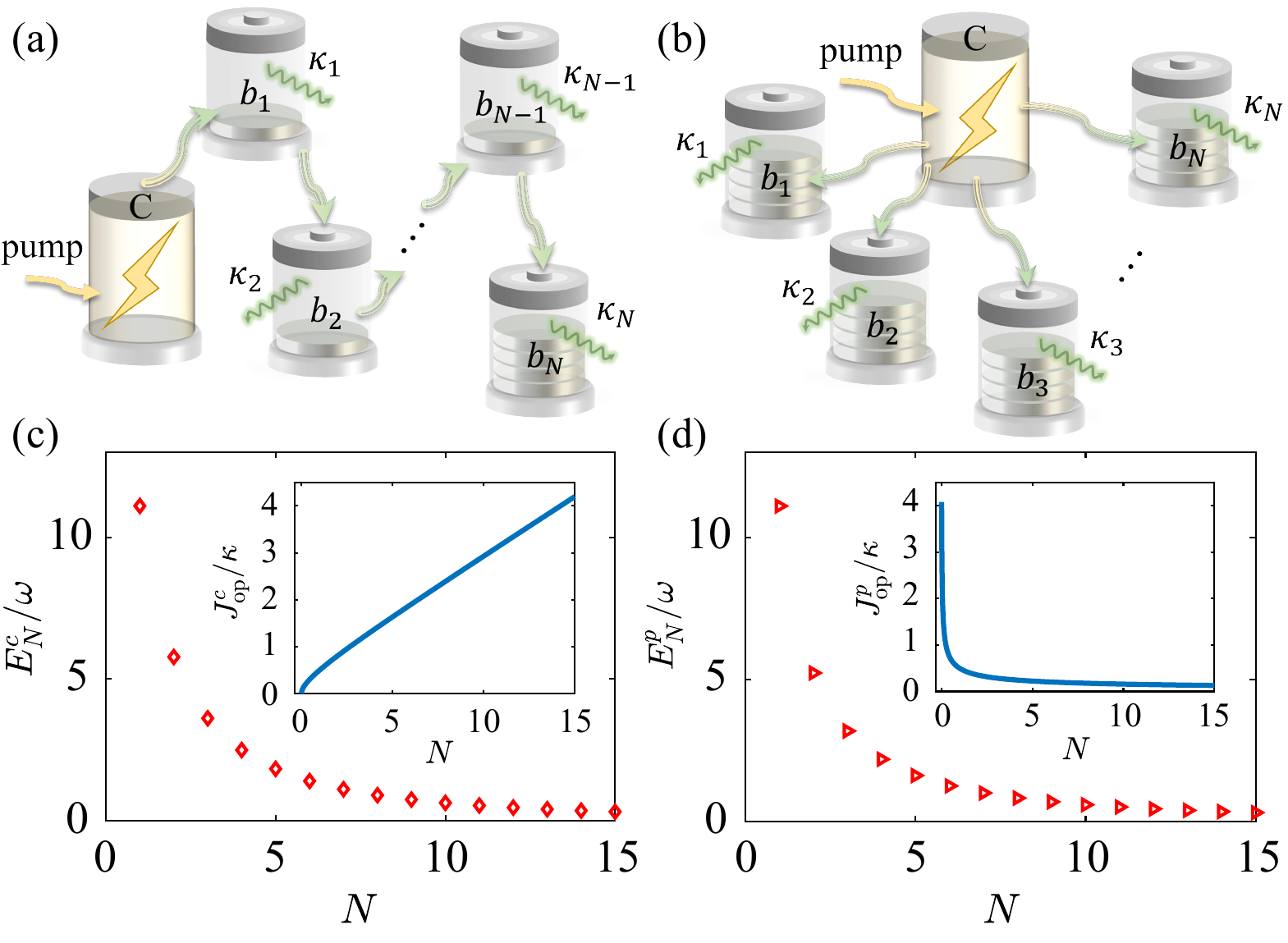}
\caption{Quantum battery networks arranged in (a) cascaded and (b) parallel configurations, consisting of a charger (C) and $N$ battery modes ($b_i$). Steady-state stored energy as a function of the battery number $N$ for (c) cascaded and (d) parallel configurations. Insets: optimal coupling strength $J_{\rm op}$ varies with $N$. The parameters used are $\kappa/\omega=0.003$, $\varepsilon/\omega=0.01$, and $J=J_{\rm{op}}$.}
         \label{fig1}
	\end{figure}

We consider a system comprising $N+1$ cavity modes with the same resonance frequency $\omega$: one charger mode and $N$ battery modes, as depicted in Fig.~\ref{fig1}. The total Hamiltonian in the interaction framework is written as $H=H_I+H_c$, where $H_I$ describes the couplings among all nodes, and $H_c=\varepsilon(a+a^{\dag})$ is the classical drive applied to the charger. We investigate two representative network architectures, namely, cascaded and parallel connection topologies. 

We first focus on the cascaded case, for which the corresponding Hamiltonian takes the form
\begin{align}
    H_I&=J_1 e^{i\theta_1} a^{\dagger}b_1+\sum_{i=2}^{N}J_ie^{i\theta_i}b_{i-1}^{\dagger}b_i+\rm{H.c.},
\end{align}
where $a$ and $b_i$ are the annihilation operators of the charger and $i$-th battery mode, respectively. The parameters $J_1$ and $J_i$ denote coherent hopping amplitudes,
and $\theta_1$ and $\theta_i$ denote the corresponding hopping phases.
    
The nonreciprocal coupling is achieved by coupling the charger $a$ and the battery $b_1$, as well as each pair $b_i$ and $b_{i+1}$, to a common reservoir~\cite{PhysRevX.5.021025,PhysRevLett.132.210402}. After adiabatic elimination of the reservoir degrees of freedom, directional energy transport emerges between different nodes of the quantum battery network. In the Markovian regime, the system dynamics is governed by the master equation
\begin{align}
\frac{d\rho}{dt}&=-i[H,\rho]+\kappa_a\mathcal{L}[a]\rho+\sum_{i=1}^N\big(\kappa_i\mathcal{L}[{b_i}]\rho+\mathcal{L}[q_i]\rho]\big).
\end{align}
The dissipative superoperator $\mathcal{L}[o]\rho=o\rho o^{\dagger}-1/2\{o^{\dagger}o,\rho\}$, where the collective operators associated with the shared reservoir are given by
$q_1=\sqrt{\Gamma_1}(p_aa+p_{b_1}b_1)$, and $q_i=\sqrt{\Gamma_i}(p_{b_{i-1}}b_{i-1}+p_{b_i}b_i)$ for $i\geq2$. Here, $\kappa_a$ and $\kappa_i$ are the local dissipation rates of the charger and battery modes into the independent baths, while $\Gamma_i$ characterizes the cooperative dissipative rate mediated by the common reservoir.

Effective nonreciprocal coupling is engineered by balancing the dissipative interactions and the coherent coupling. Unidirectional energy transfer arises when the parameters satisfy $\theta_i=\pm\pi/2$, $\mu_i=\pm1$ and $J_i=\Gamma_i/2$, where $\mu_1=p_a^*p_{b_1}$ and $\mu_i=p_{b_{i-1}}^*p_{b_i}$. The detailed derivation is presented in Appendix~\ref{AA}.
    
By solving the Heisenberg equations of motion and exploiting the recursive structure revealed by the analytical results in Appendix~\ref{AA}, we obtain an analytical expression for the energy distribution in the $n$-th battery
\begin{align}
E_n(t)=
\frac{\mathcal{A}_n}{\mathcal{V}_n^{\,2}}
\left[
\mathcal{V}_n
+
\sum_{p=0}^{n}
(-1)^{n+p}
\left(
\prod_{m\ne p}^{n}\Lambda_m
\right)
\mathcal{V}_n^{(p)}
e^{-\Lambda_p t/2}
\right]^2 ,\label{eq3}
\end{align}
where
\begin{align}
\mathcal{A}_n=
\frac{4^{\,n+1}\omega\varepsilon^2
\prod_{m=1}^{n}|\mu_m|^2\Gamma_m^2}
{\prod_{m=0}^{n}\Lambda_m^2},\notag
\end{align}
and
\begin{align}
\mathcal{V}_n &\equiv 
\prod_{0\le m<j\le n}(\Lambda_m-\Lambda_j),\notag\\
\mathcal{V}_n^{(p)} &\equiv 
\prod_{\substack{0\le m<j\le n\\ m,j\neq p}}
(\Lambda_m-\Lambda_j),\notag
\end{align}
in which $\Lambda_0=\Gamma_1|p_a|^2+\kappa_a$, $\Lambda_i=(\Gamma_i+\Gamma_{i+1})|p_{b_i}|^2+\kappa_{b_i}$, for $i=1,...,n-1$, and $\Lambda_n=\Gamma_n|p_{b_n}|^2+\kappa_{b_n}$.

In the long-time limit $t\rightarrow\infty$, dissipation drives the system to relax to a steady state, for which the stored energy of the $n$-th battery reduces to
\begin{align}
E_n^{\rm ss}=\frac{4^{\,n+1}\omega\varepsilon^2
\prod_{m=1}^{n}|\mu_m|^2\Gamma_m^2}
{\prod_{m=0}^{n}\Lambda_m^2}.
\end{align}

In the cascaded configuration, energy propagates sequentially from the charger toward the end of the chain. The terminal battery thus provides a natural figure of merit for transport performance, because it directly probes how much energy survives the cumulative attenuation and
leakage encountered along the chain. Generalizing the above results, we obtain the following dependence of the energy in the terminal battery on the battery number:
\begin{align}
E_c^{\rm{ss}}(N)=&\frac{4^{N+1}\omega\varepsilon^2\Gamma^{2N}}{(\Gamma+\kappa)^4(2\Gamma+\kappa)^{2N-2}},
\end{align}
where we have set $J_i=J$, $\Gamma_i=\Gamma$, and $\kappa_i=\kappa$ for simplicity, which is consistent with the results reported in Ref.~\cite{fbv7-m7sd}. 

For the parallel charging configuration, the interaction Hamiltonian takes the form
\begin{align}
H_I=\sum_{i=1}^{N}J_ie^{i\theta_i}a^{\dagger}b_{i}+\rm{H.c.},
\end{align}
where the cooperative dissipation via the common reservoir is characterized by the collective operators
$q_i=\sqrt{\Gamma_i}(p_aa+p_{b_i}b_i)$. The condition for nonreciprocity remains identical to that of the cascaded case, but with $\mu_i=p_ap_{b_i}$.
Solving the corresponding dynamical equations yields the energy stored in the $n$-th battery,
\begin{align}
E_n(t)=&\frac{4^{2}\omega\varepsilon^2|\mu_n|^2\Gamma_n^2}{\Lambda_0^2\Lambda_n^2(\Lambda_0-\Lambda_n)^2} \Big[(\Lambda_0-\Lambda_n)-(\Lambda_0e^{-\frac{\Lambda_n t}{2}}\notag\\
&-\Lambda_ne^{-\frac{\Lambda_0 t}{2}})\Big]^2,\label{eq7}
\end{align}	
where $\Lambda_0=\sum_{i=1}^n\Gamma_i|p_a|^2+\kappa_a$, $\Lambda_i=\Gamma_i|p_{b_i}|^2+\kappa_{b_i}$, for $i=1,...,n$. In the steady-state limit,
\begin{align}
E_n^{\rm ss}=&\frac{4^{2}\omega\varepsilon^2|\mu_n|^2\Gamma_n^2}{\Lambda_1^2\Lambda_n^2}.
\end{align}	

Similarly, according to the analytical results derived in Appendix~\ref{AA}, the steady-state stored energy as a function of the battery number $N$ can be obtained as
\begin{align}
E_p^{\rm{ss}}(N)=&\frac{4^{2}\omega\varepsilon^2\Gamma^2}{(N\Gamma+\kappa)^2(\Gamma+\kappa)^2}.
\end{align}	

The effective nonreciprocal coupling arises from the interplay between the coherent and dissipative
interactions; there exists an optimal coupling strength
$J_{\rm op}$ that maximizes the stored energy. By solving $\partial E^{\rm{ss}}/\partial J=0$, we can obtain 
\begin{align}
J_{\rm op}^c(N)=\frac{\kappa}{8}(N+\sqrt{N^2+8N}), ~~~~J_{\rm op}^p(N)=\frac{\kappa}{2\sqrt{N}},
\end{align}	
for the cascaded and parallel configurations, respectively. The corresponding maximal stored energies are shown in Fig.~\ref{fig1}. Notably, the optimal coupling exhibits distinct scalings with the number of batteries $N$ in the two configurations: it increases with transfer distance in the cascaded chain but decreases with the number of simultaneously charged batteries in the parallel geometry. This is because the cascaded topology is characterized by sequential energy propagation, whereas the parallel topology is characterized by collective energy redistribution.

In the large-$N$ limit, the optimal coupling exhibits qualitatively different scaling behaviors, with $J_{\rm op}^c\propto N$ and $J_{\rm op}^p \propto N^{-1/2}$. In the cascaded architecture, the effective propagation distance increases with the number of batteries, resulting in progressively greater energy attenuation along the chain. Consequently, a larger coupling strength is required to maintain efficient energy delivery to the terminal node. In contrast, in the parallel configuration, the charging process exhibits a collective redistribution. Increasing the coupling strength enhances not only energy injection but also loss through the cooperative dissipation channel. Consequently, the optimal coupling $J_{\rm op}^p$ decreases with $N$ due to the limitation of the nonreciprocity condition. 

\begin{figure}
	\centering
\includegraphics[width=8.6cm,height=4.88cm]{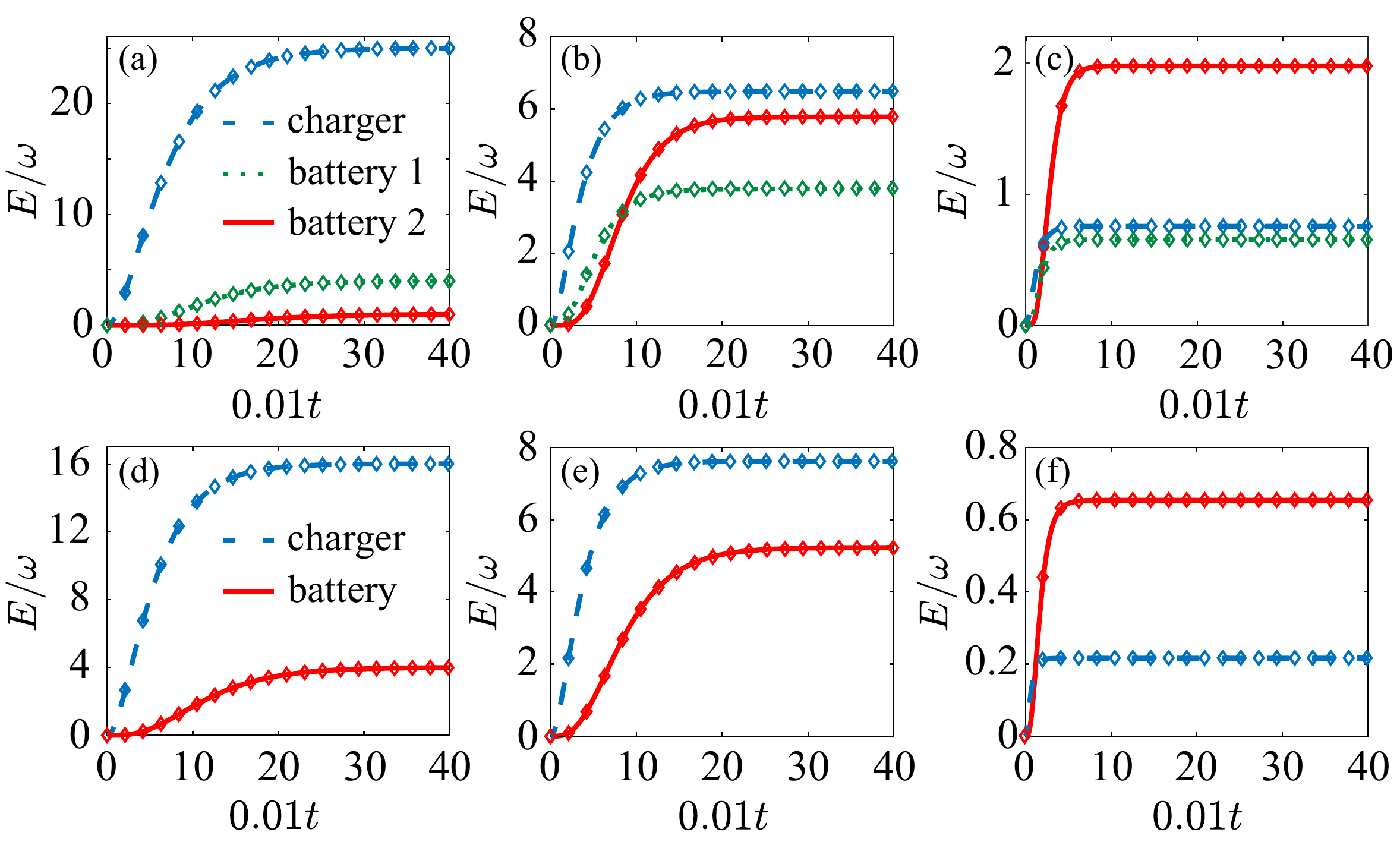}
	\caption{Energy distribution across the charger and battery modes for different coupling strengths: $J/\omega=5\times10^{-4}$~(a, d), $J=J_{\rm op}$~(b, e), and $J/\omega=10^{-2}$~(c, f). For the
parameters used here, $J^c_{\rm op}/\omega\simeq2.43\times10^{-3}$ for cascaded topology, whereas $J^p_{\rm op}/\omega\simeq1.06\times10^{-3}$ for parallel topology. Panels (a)–(c) correspond to the cascaded configuration, while panels (d)–(f) show the parallel configuration. The blue, green, and red curves represent the energies stored in the charger, the first battery, and the second battery, respectively, while the symbols represent the analytical results. In the parallel configuration, all batteries store the same amount of energy. The other parameters are $\kappa/\omega=0.003$, $\varepsilon/\omega=0.01$, and $N=2$.}
	\label{fig2}
\end{figure}

For $J\textless J_{\rm op}$, the effective nonreciprocal coupling is too weak to extract energy from the charger. In this regime, the injected energy predominantly localizes in the charger, only a small fraction is delivered to the battery. The coherent interaction is insufficient to overcome intrinsic local dissipation, thereby suppressing energy accumulation in the battery, as illustrated in Figs.~\ref{fig2}(a) and (d). Moreover, our analytical results are in excellent agreement with the numerical calculations.

As $J$ increases beyond
$J_{\rm op}$, the stronger coupling shortens the dynamical timescale, allowing the battery energy to approach its steady-state value more rapidly. However, this also amplifies cooperative (nonlocal) dissipative channels associated with the engineered nonreciprocal transmission. A substantial fraction of the injected energy is dissipated into the environment, leading to a reduced steady-state stored energy compared with the corresponding situation of $J_{\rm op}$, as shown in Fig.~\ref{fig2}. This behavior reflects a general trade-off between charging speed and energy storage performance under dissipation-engineered nonreciprocal interactions.

The optimal coupling $J_{\rm op}$ therefore corresponds to a balance between directional energy injection and environmental dissipation. At this point, nonreciprocity enables efficient energy transfer into the batteries while limiting both backflow and excessive leakage, thereby maximizing the steady-state stored energy. 

\section{Reciprocal and nonreciprocal transport mechanisms}
To obtain a more complete picture of energy transport in quantum battery networks, we now compare the reciprocal and nonreciprocal coupling cases. In contrast to directional transport, reciprocal interactions enable bidirectional energy flow and modal interference that significantly change how energy is distributed, which
spectral features dominate the response, and how the steady state is approached. 

Here, we consider the reciprocal model by removing the collective dissipative channels used to engineer nonreciprocal coupling. Thus, $\Gamma_i=0$, and the dynamics is then governed by
\begin{align}
\frac{d\rho}{dt}=-i[H,\rho]+\kappa_a\mathcal{L}[a]\rho+\sum_{i=1}^N\kappa_i\mathcal{L}[{b_i}]\rho.
\end{align} 
For the reciprocal cascaded configuration, the internal transport exhibits a pronounced odd-even effect, leading to qualitatively different scaling laws for chains with odd and even numbers of batteries. For an odd number of batteries, the terminal stored energy is \begin{align}	
E_N^{\rm ss} =\omega\left[\frac{2^{N+1}J^N\varepsilon}{\sum_{j=0}^{(N+1)/2}a(j)J^{N+1-2j}\kappa^{2j}}\right]^2,
\end{align}	
whereas for an even number of batteries, it reads 
\begin{align}
E_N^{\rm ss} =\omega\left[\frac{2^{N+1}J^N\varepsilon}{\sum_{j=0}^{N/2}b(j)J^{N-2j}\kappa^{2j+1}}\right]^2,
\end{align}	
where $a(j)$ and $b(j)$ are the constants defined in Ref.~\cite{fbv7-m7sd}.
By contrast, the reciprocal parallel configuration does not exhibit such parity-dependent transport, and its steady-state stored energy is 
\begin{align}
E_N^{\rm ss}=\frac{16J^2\omega\varepsilon^2}{(4NJ^2+\kappa^2)^2}.
\end{align}	

\begin{figure}
	\centering
	\includegraphics[width=8.5cm,height=9.754cm]{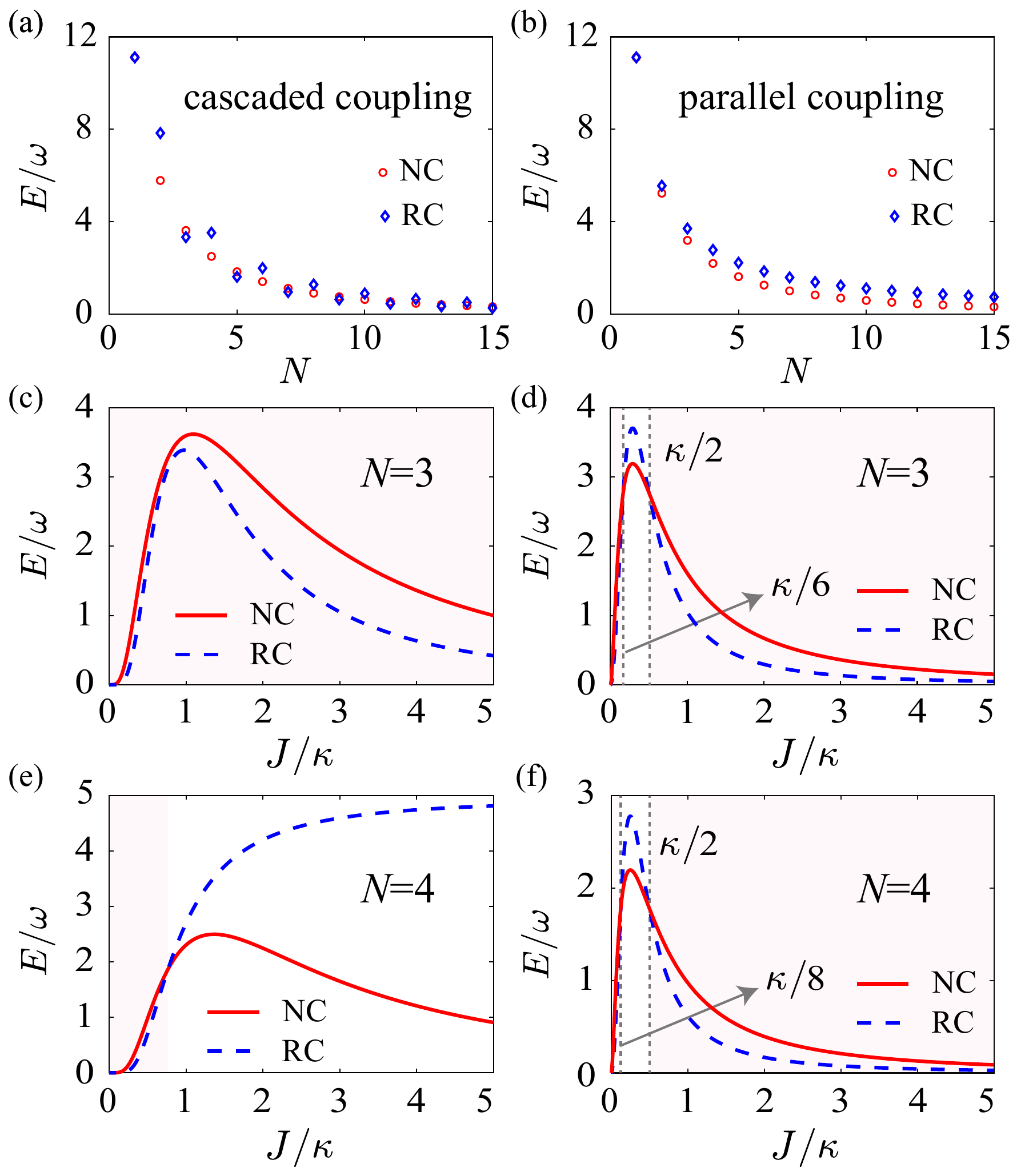}
	\caption{Steady-state stored energy as a function of the battery number $N$ under nonreciprocal coupling~(NC)~(red circles) and reciprocal coupling~(RC)~(blue diamonds) in the cascaded (a) and parallel (b) structures at $J=J_{\rm op}$. Panels (c)–(f) show the steady-state stored energy versus the coupling strength $J$: the cascaded configuration in (c, e) and the parallel configuration in (d, f), with $N=3$ for (c, d) and $N=4$ for (e, f). Solid and dashed curves are the reciprocal coupling and nonreciprocal coupling results, respectively. The other parameters are $\kappa/\omega=0.003$, $\varepsilon/\omega=0.01$.}
	\label{fig3}
\end{figure}

At the optimal coupling $J_{\rm op}$ mentioned above, the steady-state stored energy exhibits distinct behaviors for reciprocal and nonreciprocal interactions, as illustrated in Figs.~\ref{fig3}(a) and \ref{fig3}(b). For the reciprocal cascaded coupling, the steady-state energy can be understood from the spectral structure of the underlying tight-binding Hamiltonian, which can be expressed in terms of the Green function, whose modal contribution scales as $(-E_k+i\kappa/2)^{-1}$,~see Appendix~\ref{AB} for a detailed discussion. As a result, the zero-energy eigenmode acquires greater weight and therefore dominates the steady-state energy distribution. For the charging process of an even number of batteries, there always exists a zero-energy eigenmode, which makes a dominant contribution to the steady-state amplitude, enabling constructive energy transport from the charger to the end battery. In contrast, for an odd number of batteries, the spectrum is symmetric around zero energy, and no zero-energy eigenmode exists. Consequently, the contributions from different transport modes interfere destructively, suppressing the steady-state energy and resulting in the observed odd–even effect.

In addition, this spectral mechanism also determines the dependence of stored energy on the coupling strength. When $N$ is odd, the stored energy exhibits a pronounced peak as a function of the coupling strength, as shown in Fig.~\ref{fig3}(c). In the weak-coupling regime, energy transfer from the charger to the batteries is inefficient, resulting in low stored energy. When the coupling strength becomes sufficiently large, the charger and the batteries hybridize into normal modes whose frequencies mismatch with the driving field. This detuning reduces charging efficiency and consequently decreases stored energy.

On the contrary, when $N$ is even, the system supports a zero-energy mode that remains resonant with the driving field. As a result, there is no frequency mismatch induced by strong coupling. In this case, the stored energy increases with the coupling strength and ultimately approaches a saturation value, without decreasing, as shown in Fig.~\ref{fig3}(e).

\begin{figure}
	\centering
	\includegraphics[width=8.5cm,height=5cm]{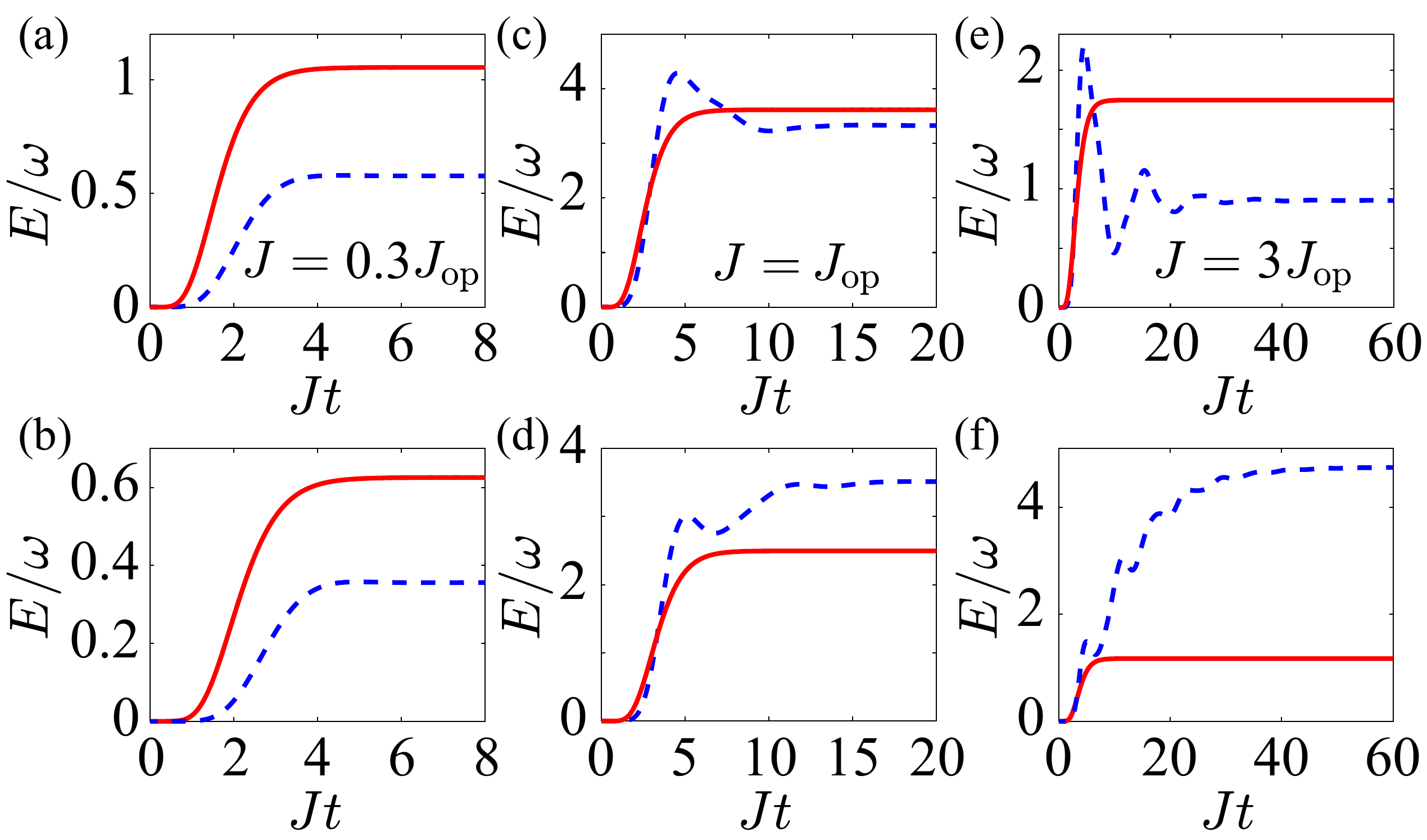}
	\caption{Time evolution of the stored energy in the cascaded configuration for different coupling strengths: $J=0.3J_{\rm op}$~(a, b), $J=J_{\rm op}$~(c, d), and $J=3J_{\rm op}$~(e, f). The top row corresponds to the case of $N=3$, while the bottom row corresponds to $N=4$. Solid and dashed curves are the reciprocal coupling and nonreciprocal coupling results, respectively. The other parameters are $\kappa/\omega=0.003$, $\varepsilon/\omega=0.01$.}
    \label{fig4}
\end{figure}

In contrast to the reciprocal systems, nonreciprocal coupling suppresses energy backflow; there always exists an optimal coupling $J_{\rm op}$. For the cascaded odd-$N$ chains shown in Fig.~\ref{fig3}(c), the nonreciprocal curve
lies above the reciprocal one. For the even-$N$, however, the reciprocal system can
outperform the nonreciprocal system because of the emergence of the zero-energy eigenmode, as shown in Fig.~\ref{fig3}(e). The nonreciprocal advantage is then restricted to the small $J$ regime. From the crossing points extracted from the numerical
curves, we find an approximate boundary $J/\kappa=0.1156N+0.2972$, which increases with the number of batteries.

For the parallel configuration, no mechanism analogous to the zero-energy-mode enhancement in the cascaded configuration exists. All zero-energy dark states are orthogonal to the charging vector and therefore remain decoupled from the charging dynamics. Consequently, the system dynamics is governed solely by the two bright states (see Appendix~\ref{ACcc}), leading to a much simpler Green-function expansion and steady-state response. In the strong-coupling regime, the hybridized modes are also shifted away from the driving frequency, reducing the stored energy. We identify a parameter window
\begin{equation}
\frac{\kappa}{2N} < J < \frac{\kappa}{2},
\end{equation}
within which reciprocal coupling can yield larger stored energy than nonreciprocal
coupling. This is because reservoir-engineered nonreciprocal interaction is accompanied by an additional collective dissipation channel, whereas the reciprocal configuration involves energy backflow without such an extra dissipative process. Within this window, the reduction in energy backflow provided by the nonreciprocal coupling is insufficient to compensate for the accompanying reservoir-induced dissipation. Consequently, the reciprocal configuration exhibits a slightly larger steady-state stored energy than the nonreciprocal one. This regime, however, is relatively narrow, as shown in Figs.~\ref{fig3}(d) and (f). Outside this window, the directional nonreciprocal architecture becomes advantageous because it suppresses the backflow of energy.

\begin{figure}
	\centering
	\includegraphics[width=8.5cm,height=5.02cm]{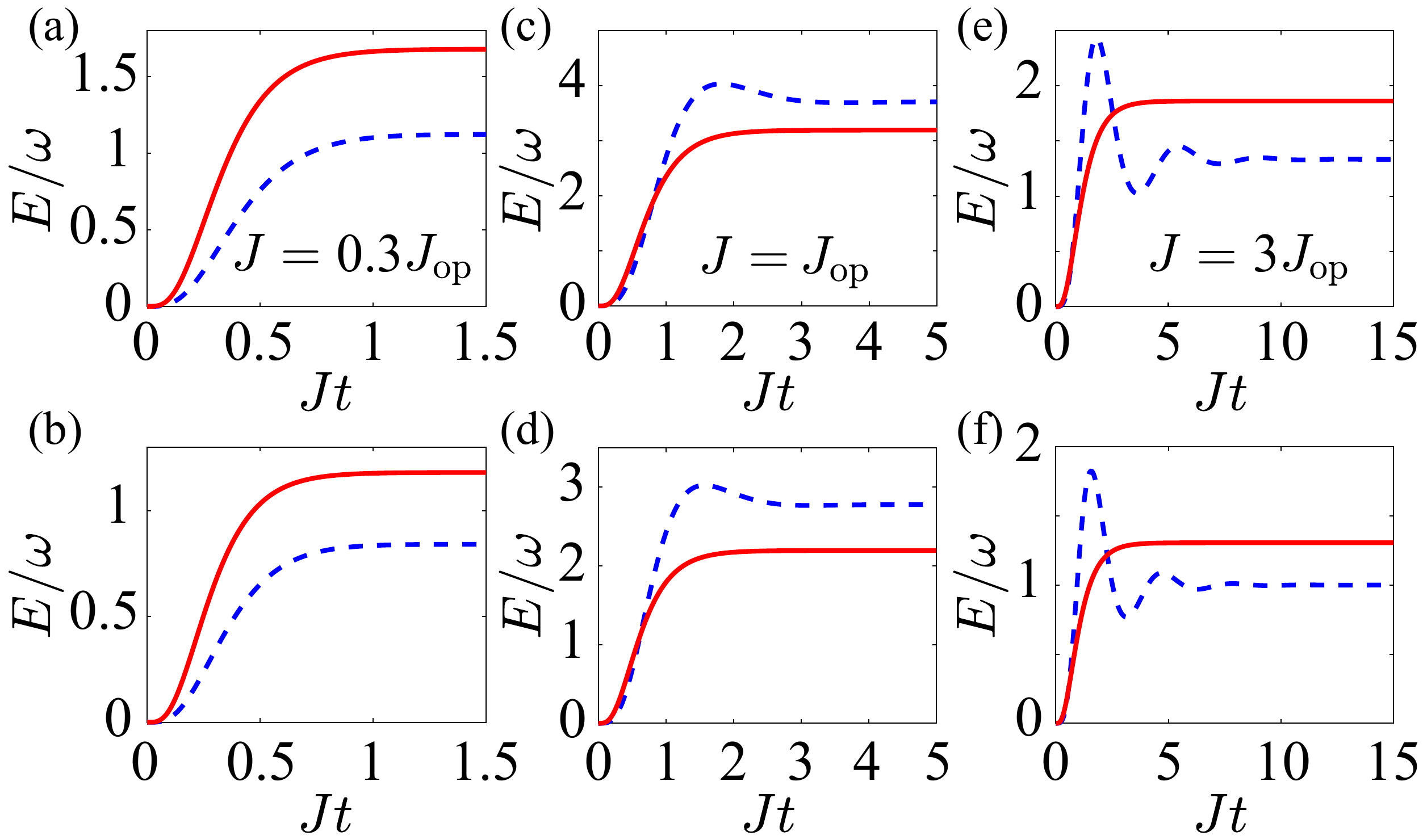}
	\caption{Time evolution of the stored energy in the parallel configuration for different coupling strengths: $J=0.3J_{\rm op}$~(a, b), $J=J_{\rm op}$~(c, d), and $J=3J_{\rm op}$~(e, f). The top row corresponds to the case of $N=3$, while the bottom row corresponds to $N=4$. Solid and dashed curves are the reciprocal coupling and nonreciprocal coupling results, respectively. The other parameters are $\kappa/\omega=0.003$, $\varepsilon/\omega=0.01$.}
    \label{fig51}
\end{figure}

We further compare the charging dynamics for different coupling configurations, as shown in Figs.~\ref{fig4} and \ref{fig51}. The steady-state energy quantifies the amount
of energy eventually stored, but it does not specify how rapidly the network reaches its operating point. In the weak-coupling regime, the energy is predominantly dissipated into the environment, and no energy oscillation is observed in the reciprocal coupling case. For both cascaded and parallel configurations, the nonreciprocal coupling consistently yields higher stored energy than the reciprocal case. The charging power is quantified by $P=E/t$. Consistent with the results reported in Ref.~\cite{fbv7-m7sd}, the nonreciprocal coupling yields a larger charging power. As $J$ increases, distinct transport behaviors emerge in the two configurations. In the reciprocal cascaded configuration, energy propagates sequentially along the chain. Oscillatory exchange between neighboring nodes slows down the net energy transfer, resulting in a longer relaxation time that increases approximately linearly with the number of batteries $N$. In contrast, in the parallel configuration, the charger injects energy into all batteries simultaneously, making the relaxation time much shorter and independent of $N$.

The situation changes qualitatively for nonreciprocal coupling. In this case, energy is delivered monotonically from the charger to the batteries, the dwell time in intermediate nodes is reduced, and the relaxation toward the steady state is accelerated. For larger coupling strengths ($J\textgreater J_{\rm op}$), apart from the enhancement of the reciprocal coupling induced by the zero-energy eigenmode in the even-$N$ case, the nonreciprocal configurations generally exhibit higher stored energy and charging power. Although the reciprocal coupling may exhibit a larger first oscillation peak, such an enhancement is only transient.

\section{Reservoir consideration}
\begin{figure}
	\centering
\includegraphics[width=8.5cm,height=9.063cm]{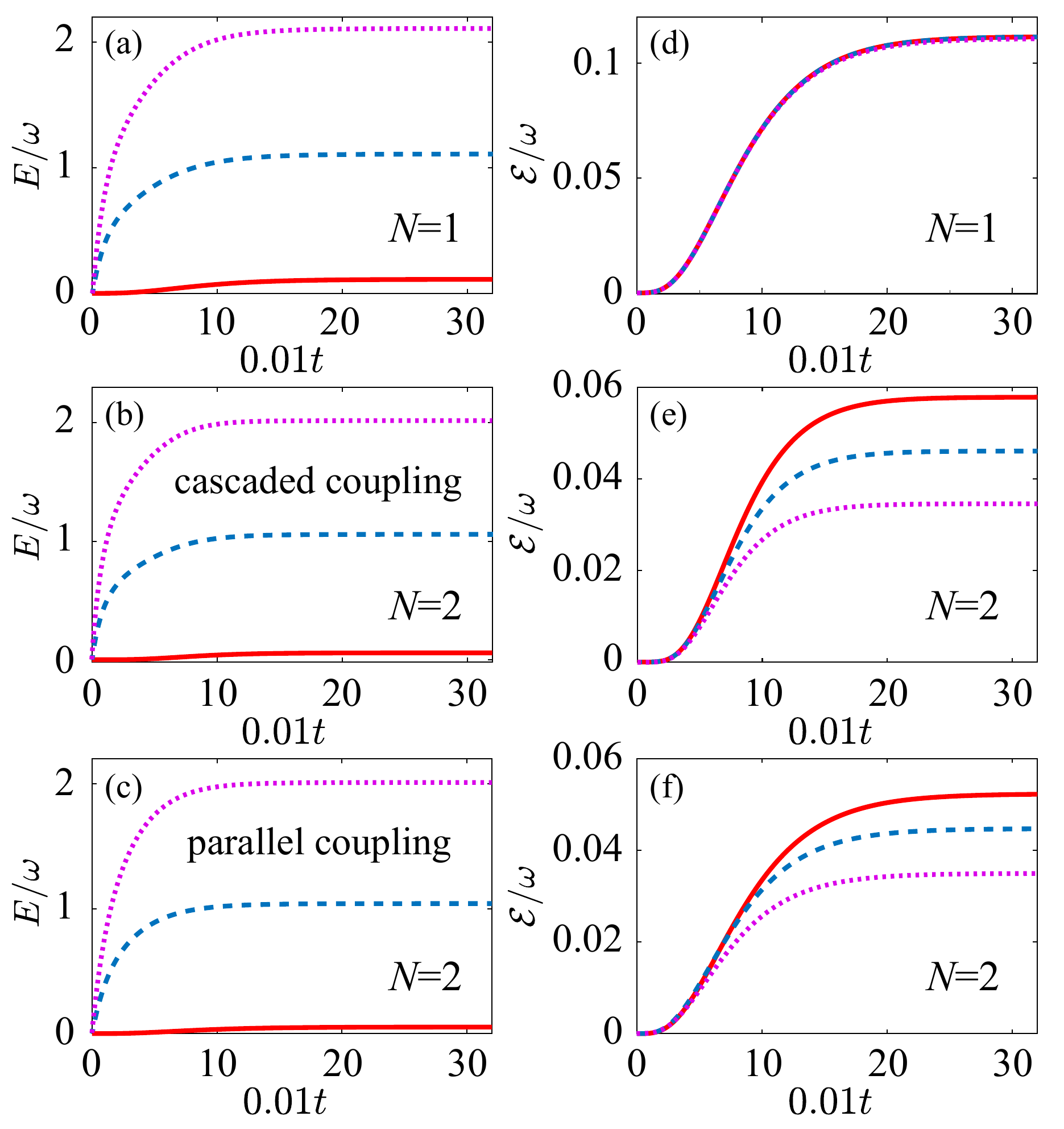}
	\caption{Dynamics of stored energy and ergotropy. Top row: single battery~($N=1$). Middle row: cascaded configuration with  $N=2$. Bottom row: parallel configuration with $N=2$. The solid line, dashed line, and dotted line represent $n_{\rm th}=0$, $n_{\rm th}=1$, and $n_{\rm th}=2$, respectively. The other parameters are $\kappa/\omega=0.003$, $\varepsilon/\omega=0.001$, and $J=J_{\rm op}$.}
    \label{fig5}
\end{figure}

Because the nonreciprocal coupling is mediated by a common reservoir, it is important to examine how its properties affect the charging performance. We first consider a thermal environment, for which the system dynamics is governed by the master equation
\begin{align}
\dot{\rho}&=-i[H,\rho]+n_{\rm th}\Big[\kappa_a\mathcal{L}[a^{\dag}]\rho+\sum_{i=1}^N(\kappa_i\mathcal{L}[b^{\dag}_i]\rho+\mathcal{L}[q^{\dag}_i]\rho)\Big] \notag\\
&~~~~~+(n_{\rm th}+1)\Big[\kappa_a\mathcal{L}[a]\rho+\sum_{i=1}^N(\kappa_i\mathcal{L}[b_i]\rho+\mathcal{L}[q_i]\rho)\Big],
\end{align}
where $n_{\rm th}$ is the average number of photons in the reservoir. 
As illustrated in Fig.~\ref{fig5}, we show the time evolution of the stored energy in the quantum battery. In the thermal bath, environment-induced excitations are continuously injected into the cavity modes via Lindblad processes. The fluctuation strength associated with the thermal reservoir scales linearly with the bath occupation, the induced incoherent pumping of photons into the system is correspondingly enhanced, resulting in a steady-state energy that increases linearly with $n_{\rm th}$, which can be expressed as $$E^{\rm ss'}_{c(p)}=E^{\rm ss}_{c(p)}+n_{\rm th},$$ for both cascaded and parallel topologies. We note that, due to the larger photon number and rapid growth of the Hilbert-space dimension with the battery number $N$, our simulations are performed at a relatively weak drive to ensure convergence in a truncated Fock space.

However, not all of the stored energy can be utilized; ergotropy (extractable work), defined as $\mathcal{E}(t)=\mathrm{Tr}[H_B\rho_B(t)]-\mathop{\min}\limits_{U}\mathrm{Tr}[H_BU\rho_B(t)U^{\dag}]$, is used to quantify the performance of the battery~\cite{RevModPhys.96.031001,Allahverdyan_2004,PhysRevLett.133.180401}, where the subscript ``$B$" denotes the battery subsystem and $H_B=\omega\sum_{i=1}^Nb_i^{\dag}b_i$, $\rho_B$ is the corresponding density matrix. The battery is initially prepared in ground state, the system remains in a pure state throughout the evolution process for a zero-temperature vacuum reservoir. As a result, the stored energy is fully nonpassive and equals the ergotropy $\mathcal{E}$ ~\cite{PhysRevA.107.042419,PhysRevA.109.042207}. In contrast, hot reservoir drives the system toward mixed states due to thermal fluctuations. Although thermal excitations can increase the total population $\langle b^{\dagger}b\rangle$, they mainly contribute disordered energy that cannot be efficiently converted into useful work. As a result, the enhancement of stored energy does not imply improved battery performance, and the ergotropy may even be reduced in the presence of thermal noise.

As shown in Figs.~\ref{fig5}(d-f), the ergotropy in the hot reservoir is lower than that in the vacuum reservoir, and decreases as the reservoir temperature increases. This behavior originates from the fact that thermal fluctuations primarily inject incoherent excitations into the system. These results indicate that, while thermal environments can enhance overall energy storage, low-temperature reservoirs are more favorable for preserving coherent energy accumulation and maintaining a larger extractable fraction, highlighting the distinction between energy storage and useful work extraction in quantum batteries.
\begin{figure}
	\centering
\includegraphics[width=8.5cm,height=6.84cm]{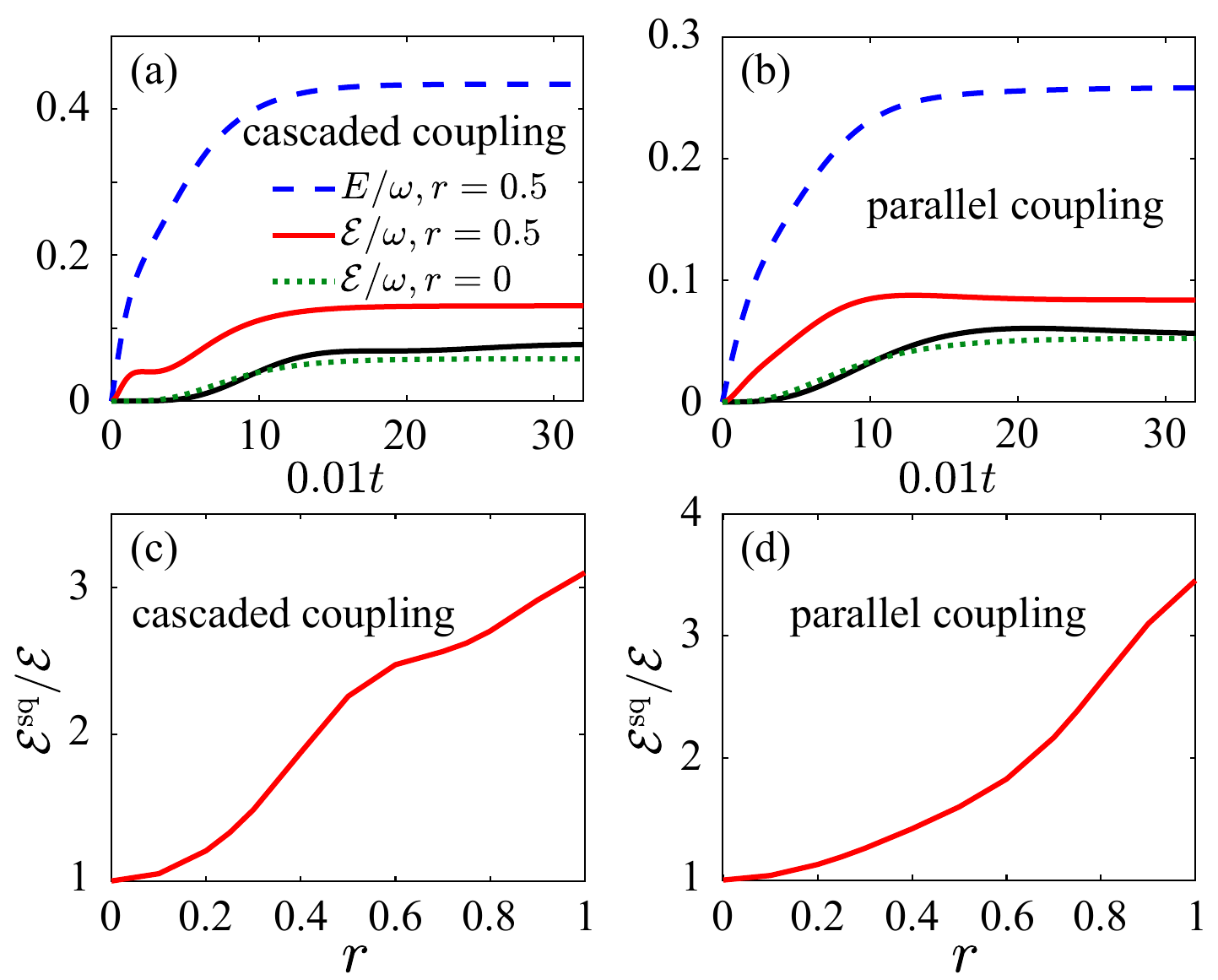}
	\caption{Evolution of the stored energy and ergotropy under nonreciprocal coupling for the cascaded~(a) and parallel~(b) configurations. Black solid curves correspond to reciprocal coupling for comparison.  Enhancement factor of the ergotropy, $\mathcal{E}^{\rm sq}/\mathcal{E}$, as a function of the squeezing parameter $r$ for the cascaded (c) and parallel (d) configurations. Here, $\mathcal{E}^{\rm sq}$ denotes the ergotropy in the presence of squeezing, while $\mathcal{E}$ corresponds to the case without squeezing. The parameters are $\kappa/\omega=0.003$, $\varepsilon/\omega=0.001$, $J=J_{\rm op}$, and $N=2$.}
    \label{fig6}
\end{figure}

We further consider the effect of a squeezed reservoir. The dynamics of the system is described by
\begin{align}
\dot{\rho}&=-i[H,\rho]+P\Big[\kappa_a\mathcal{L}[a^{\dag}]\rho+\sum_{i=1}^N(\kappa_i\mathcal{L}[b^{\dag}_i]\rho+\mathcal{L}[q^{\dag}_i]\rho)\Big]\notag\\
&~~~~
+(P+1)\Big[\kappa_a\mathcal{L}[a]\rho+\sum_{i=1}^N(\kappa_i\mathcal{L}[b_i]\rho+\mathcal{L}[q_i]\rho)\Big]\notag\\
&~~~~-Q\Big[\kappa_a\mathcal{S}[a]\rho+\sum_{i=1}^N(\kappa_i\mathcal{S}[b_i]\rho+\mathcal{S}[q_i]\rho)\Big]\notag\\
&~~~~-Q^*\Big[\kappa_a\mathcal{S}[a^{\dag}]\rho+\sum_{i=1}^N(\kappa_i\mathcal{S}[b^{\dag}_i]\rho+\mathcal{S}[q^{\dag}_i]\rho)\Big],
\end{align}
where $\mathcal{S}[o]\rho=o\rho o-1/2\{oo,\rho\}$, $P=\sinh^2{r}$, and $Q=\sinh{r}\cosh{r}e^{-i\theta_r}$, with $r$ and $\theta_r$ denote the squeezing strength and phase. Unlike the thermal bath, the squeezed reservoir not only provides excitations but also introduces quantum correlations through two-photon processes. These correlations modify the fluctuation properties of the environment and convert environmental fluctuations into ordered energy that can be extracted through unitary operations. As illustrated in Fig.~\ref{fig6}, the squeezed reservoir enhances the ergotropy of the quantum battery. Moreover, the extractable work increases with the squeezing strength, demonstrating that stronger squeezing not only increases the stored energy but also enhances the useful fraction.
\section{Design principles and feasible implementation}
Our work extends single-cell quantum batteries by incorporating spatial transport effects and establishes transport-oriented design rules for battery networks.
First, topology determines the main transport constraint: cascaded chains exhibit sequential energy propagation, whereas parallel networks exhibit collective energy redistribution. Second, coupling directionality controls energy backflow and modal interference, making nonreciprocal interactions particularly advantageous for fast relaxation. Third, reservoir properties determine whether the transported energy remains passive or becomes extractable work: thermal noise mainly raises passive energy, whereas squeezing enhances nonpassive energy and ergotropy.

Arrays of coupled optical or microwave cavities provide a natural platform for implementing the proposed architectures, with both coherent hopping and engineered dissipation being precisely controllable. Collective dissipation can be engineered via strongly damped auxiliary cavity modes or a waveguide. Adiabatic elimination of these auxiliary modes yields an effective dissipative coupling, enabling directional energy transport. This offers a feasible route toward realizing nonreciprocal quantum battery networks using existing photonic and superconducting cavity technologies.

\section{Conclusion}
In conclusion, we have studied how topology, nonreciprocal coupling, and different reservoirs affect energy transport in quantum battery networks. For cascaded and parallel architectures, we derived analytical results that reveal distinct topology-dependent optimal-coupling laws, reflecting the contrast between sequential-delivery transport and collective-redistribution charging. We further showed that reciprocal cascaded networks exhibit a parity-dependent spectral transport mechanism, which explains the odd-even response and is absent in the nonreciprocal
and parallel configurations. Finally, by comparing thermal and squeezed reservoirs, we demonstrated that stored energy and ergotropy can exhibit qualitatively different responses to the environment.
These results highlight the importance of transport engineering and work extraction and provide a basic framework for the design of quantum battery networks.

\begin{acknowledgments}
This work was supported by the National Natural Science Foundation of China (Grants No. 12274376, No. 12575032, No. 12125406, No. U24A2015). 
\end{acknowledgments}
\section*{ DATA AVAILABILITY}
	The data that support the findings of this article are openly available~\cite{liu_2026_19182043}.
\appendix
\section{Analytical solution of the charging dynamics for cascaded and parallel configurations}\label{AA}
We first consider a minimal system to illustrate the analytical procedure; this structure enables the analytical results to be extended straightforwardly to an arbitrary number of batteries.
\subsection{Cascaded configuration}
For the cascaded coupling with $N=2$, the equations of motion for the first moments read
\begin{align}
\frac{d\langle a\rangle}{dt}=&-\frac{\Lambda_0}{2}\langle a\rangle-(iJ_1e^{i\theta_1}+\frac{\mu_1\Gamma_1}{2})\langle b_1\rangle-i\varepsilon,\notag\\
\frac{d\langle b_1\rangle}{dt}=&-\frac{\Lambda_1}{2}\langle b_1\rangle-(iJ_1e^{-i\theta_1}+\frac{\mu_1^*\Gamma_1}{2})\langle a\rangle-(iJ_2e^{i\theta_2}\notag\\&
+\frac{\mu_2\Gamma_2}{2})\langle b_2\rangle,\notag\\
\frac{d\langle b_2\rangle}{dt}=&-\frac{\Lambda_2}{2}\langle b_2\rangle-(iJ_2e^{-i\theta_2}+\frac{\mu_2^*\Gamma_2}{2})\langle b_1\rangle,
\end{align}
where $\mu_1=p_a^*p_{b1}$, $\mu_2=p_{b_1}^*p_{b_2}$, and $\Lambda_0=\Gamma_1|p_a|^2+\kappa_a$, $\Lambda_1=(\Gamma_1+\Gamma_2)|p_{b_1}|^2+\kappa_{b_1}$, $\Lambda_2=\Gamma_2|p_{b_2}|^2+\kappa_{b_2}$. Nonreciprocal coupling can be realized when $\theta_1=\theta_2=\pm\pi/2$, $p_a=\pm1$, $p_{b_1}=\pm1$, $p_{b_2}=\pm1$, or $\theta_1=\theta_2=0$, $p_a=i$, $p_{b_1}=1$, $p_{b_2}=-i$, and $J_i=\Gamma_i/2$. It is worth noting that the complex-valued relative dissipative coupling strengths, $p_a$, $p_{b_1}$, and $p_{b_2}$, considered here cannot be directly realized in waveguide-based experiments. Therefore, these parameters are not considered in the following analysis.

Solving the above linear equations yields the stored energies
\begin{align}
E_a(t)=&|\langle a\rangle|^2=\frac{4\omega\varepsilon^2(1-e^{-\frac{1}{2}t\Lambda_0})^2}{\Lambda_0^2},\notag\\
E_{b_1}(t)=&|\langle b_1\rangle|^2=\frac{16\omega\varepsilon^2|\mu_1|^2\Gamma_1^2}{\Lambda_0^2\Lambda_1^2(\Lambda_0-\Lambda_1)^2}\Big[(\Lambda_0-\Lambda_1)\notag\\
&-(\Lambda_0 e^{-\frac{1}{2}\Lambda_1 t} -\Lambda_1 e^{-\frac{1}{2}\Lambda_0 t})\Big]^2,\notag\\
E_{b_2}(t)=&|\langle b_2\rangle|^2\notag\\
=&\frac{64\omega\varepsilon^2|\mu_1|^2|\mu_2|^2\Gamma_1^2\Gamma_2^2}{\Lambda_0^2\Lambda_1^2\Lambda_2^2(\Lambda_0-\Lambda_1)^2(\Lambda_1-\Lambda_2)^2(\Lambda_0-\Lambda_2)^2}\notag\\
&\times\Big[(\Lambda_0-\Lambda_1)(\Lambda_1-\Lambda_2)(\Lambda_0-\Lambda_2)-\Lambda_0\Lambda_1\notag\\
&\times (\Lambda_0-\Lambda_1)e^{-\frac{1}{2}\Lambda_2t}+\Lambda_0\Lambda_2(\Lambda_0-\Lambda_2)e^{-\frac{1}{2}\Lambda_1t}\notag\\
&-\Lambda_1\Lambda_2(\Lambda_1-\Lambda_2)e^{-\frac{1}{2}\Lambda_0t}\Big]^2.
\end{align}

For $N=3$, the equations of motion are given by
\begin{align}
	\frac{d\langle a\rangle}{dt}=&-\frac{\Lambda_0}{2}\langle a\rangle-\big(iJ_1e^{i\theta_1}+\frac{\mu_1\Gamma_1}{2}\big)\langle b_1\rangle-i\varepsilon\notag
	\\
	\frac{d\langle b_1\rangle}{dt}=&-\frac{\Lambda_1}{2}\langle b_1\rangle-\big(iJ_1e^{-i\theta_1}+\frac{\mu_1^*\Gamma_1}{2}\big)\langle a\rangle-(iJ_2e^{i\theta_2}\notag\\
    &+\frac{\mu_2\Gamma_2}{2})\langle b_2\rangle\notag
	\\
	\frac{d\langle b_2\rangle}{dt}&=-\frac{\Lambda_2}{2}\langle b_2\rangle-\big(iJ_2e^{-i\theta_2}+\frac{\mu_2^*\Gamma_2}{2}\big)\langle b_1\rangle-(iJ_3e^{i\theta_3}\notag
	\\
    &+\frac{\mu_3\Gamma_3}{2})\langle b_3\rangle\notag
	\\
	\frac{d\langle b_3\rangle}{dt}&=-\frac{\Lambda_3}{2}\langle b_3\rangle-\big(iJ_3e^{-i\theta_3}+\frac{\mu_3^*\Gamma_3}{2}\big)\langle b_2\rangle,
\end{align}
where $\mu_1=p_a^*p_b$, $\mu_2=p_{b_1}^*p_{b_2}$, $\mu_3=p_{b_2}^*p_{b_3}$, and $\Lambda_0=\Gamma_1|p_a|^2+\kappa_a$, $\Lambda_1=(\Gamma_1+\Gamma_2)|p_{b_1}|^2+\kappa_{b_1}$, $\Lambda_2=(\Gamma_2+\Gamma_3)|p_{b_2}|^2+\kappa_{b_2}$, $\Lambda_3=\Gamma_3|p_{b_3}|^2+\kappa_{b_3}$. Nonreciprocal coupling can be realized by setting $\theta_1=\theta_2=\theta_3=\pm\pi/2$, $p_a=p_{b_1}=p_{b_2}=p_{b_3}=\pm1$, and $J_i=\Gamma_i/2$.  

The corresponding stored energies read
\begin{align}
E_a(t)=&|\langle a\rangle|^2=\frac{4\omega\varepsilon^2(1-e^{-\frac{1}{2}t\Lambda_0})^2}{\Lambda_0^2},\notag\\
E_{b_1}(t)=&|\langle b_1\rangle|^2=\frac{16\omega\varepsilon^2|\mu_1|^2\Gamma_1^2}{\Lambda_0^2\Lambda_1^2(\Lambda_0-\Lambda_1)^2}\Big[(\Lambda_0-\Lambda_1)\notag\\
&-(\Lambda_0 e^{-\frac{1}{2}\Lambda_1 t} -\Lambda_1 e^{-\frac{1}{2}\Lambda_0 t})\Big]^2,\notag\\
E_{b_2}(t)=&|\langle b_2\rangle|^2\notag\\
=&\frac{64\omega\varepsilon^2|\mu_1|^2|\mu_2|^2\Gamma_1^2\Gamma_2^2}{\Lambda_0^2\Lambda_1^2\Lambda_2^2(\Lambda_0-\Lambda_1)^2(\Lambda_1-\Lambda_2)^2(\Lambda_0-\Lambda_2)^2}\notag\\
&\Big[(\Lambda_0-\Lambda_1)(\Lambda_1-\Lambda_2)(\Lambda_0-\Lambda_2)-\Lambda_0\Lambda_1\notag\\
&\times (\Lambda_0-\Lambda_1)e^{-\frac{1}{2}\Lambda_2t}+\Lambda_0\Lambda_2(\Lambda_0-\Lambda_2)e^{-\frac{1}{2}\Lambda_1t}\notag\\
&-\Lambda_1\Lambda_2(\Lambda_1-\Lambda_2)e^{-\frac{1}{2}\Lambda_0t}\Big]^2.\notag\\
E_{b_3}(t)
=&\,|\langle b_3\rangle|^2
\notag\\
=&\;
\frac{256\omega\varepsilon^2
|\mu_1|^2|\mu_2|^2|\mu_3|^2
\Gamma_1^2\Gamma_2^2\Gamma_3^2}
{\Lambda_0^2\Lambda_1^2\Lambda_2^2\Lambda_3^2
(\Lambda_0-\Lambda_1)^2(\Lambda_0-\Lambda_2)^2(\Lambda_0-\Lambda_3)^2}
\notag\\
&\times
\frac{1}
{(\Lambda_1-\Lambda_2)^2(\Lambda_1-\Lambda_3)^2(\Lambda_2-\Lambda_3)^2}
\notag\\[4pt]
&\times
\Big[
(\Lambda_0-\Lambda_1)(\Lambda_0-\Lambda_2)(\Lambda_0-\Lambda_3)(\Lambda_1-\Lambda_2)
\notag\\
&\times
(\Lambda_1-\Lambda_3)(\Lambda_2-\Lambda_3)
\notag-\Lambda_0\Lambda_1\Lambda_2
(\Lambda_0-\Lambda_1)\\[3pt]
&
\times(\Lambda_0-\Lambda_2)(\Lambda_1-\Lambda_2)
e^{-\frac{\Lambda_3 t}{2}}+\Lambda_0\Lambda_1\Lambda_3
(\Lambda_0
\notag\\
&
-\Lambda_1)(\Lambda_0-\Lambda_3)(\Lambda_1-\Lambda_3)
e^{-\frac{\Lambda_2 t}{2}}-\Lambda_0\Lambda_2\Lambda_3
\notag\\
&
\times(\Lambda_0-\Lambda_2)(\Lambda_0-\Lambda_3)(\Lambda_2-\Lambda_3)
e^{-\frac{\Lambda_1 t}{2}}+\Lambda_1
\notag\\
&
\times\Lambda_2\Lambda_3
(\Lambda_1-\Lambda_2)(\Lambda_1-\Lambda_3)(\Lambda_2-\Lambda_3)
e^{-\frac{\Lambda_0 t}{2}}
\Big]^2 .
\end{align}
The analytical expressions for the $N\geq4$ cases are omitted here due to their lengthy form. By analyzing these results, one can identify the underlying recursive structure and derive the general energy distribution for arbitrary $n$, leading to Eq.~(\ref{eq3}) in the main text.
\subsection{Parallel configuration}
In the parallel configuration, the charger couples to all battery modes simultaneously. For $N=2$, the operator equations become
\begin{align}
 	\frac{d\langle a\rangle}{dt}=&-\frac{\Lambda_0}{2}\langle a\rangle-\big(iJ_1e^{i\theta_1}+\frac{\mu_1\Gamma_1}{2}\big)\langle b_1\rangle\notag
	\\&-\big(iJ_2e^{i\theta_2}+\frac{\mu_2\Gamma_2}{2}\big)\langle b_2\rangle-i\varepsilon\notag
	\\
	\frac{d\langle b_1\rangle}{dt}=&-\frac{\Lambda_1}{2}\langle b_1\rangle-\big(iJ_1e^{-i\theta_1}+\frac{\mu_1^*\Gamma_1}{2}\big)\langle a\rangle\notag
	\\
	\frac{d\langle b_2\rangle}{dt}=&-\frac{\Lambda_2}{2}\langle b_2\rangle-\big(iJ_2e^{-i\theta_2}+\frac{\mu_2^*\Gamma_2}{2}\big)\langle a\rangle   
\end{align}
where $\mu_1=p_a^*p_{b_1}$, $\mu_2=p_a^*p_{b_2}$ and $\Lambda_0=\kappa_{a}+(\Gamma_{1}+\Gamma_{2})|p_a|^2$, $\Lambda_1=\kappa_{b_1}+\Gamma_{1}|p_{b_1}|^2$ and $\Lambda_2=\kappa_{b_2}+\Gamma_{2}|p_{b_2}|^2$. Nonreciprocal coupling can be realized when $\theta_1=\theta_2=\pm\pi/2$, $p_a=p_{b_1}=p_{b_2}=\pm1$, and $J_i=\Gamma_i/2$. 

The corresponding stored energies are
\begin{align}
	E_a(t)=&\frac{4\omega\varepsilon^2(1-e^{-\frac{1}{2}t\Lambda_0})^2}{\Lambda_0^2},\notag\\  E_{b_1}(t)=&\frac{16\omega\varepsilon^2|\mu_1|^2\Gamma_1^2}{\Lambda_0^2\Lambda_1^2(\Lambda_0-\Lambda_1)^2}\Big[(\Lambda_0-\Lambda_1)-(\Lambda_0 e^{-\frac{1}{2}\Lambda_1 t}\notag\\
    &-\Lambda_1 e^{-\frac{1}{2}\Lambda_0 t})\Big]^2,\notag\\  E_{b_2}(t)=&\frac{16\omega\varepsilon^2|\mu_2|^2\Gamma_2^2}{\Lambda_0^2\Lambda_2^2(\Lambda_0-\Lambda_2)^2}\Big[(\Lambda_0-\Lambda_2)-(\Lambda_0 e^{-\frac{1}{2}\Lambda_2 t}\notag\\
    &-\Lambda_2 e^{-\frac{1}{2}\Lambda_0 t})\Big]^2.
\end{align}

For $N=3$, the equations of motion are
\begin{align}
 	\frac{d\langle a\rangle}{dt}=&-\frac{\Lambda_0}{2}\langle a\rangle-\big(iJ_1e^{i\theta_1}+\frac{\mu_1\Gamma_1}{2}\big)\langle b_1\rangle\notag
	\\&-\big(iJ_2e^{i\theta_2}+\frac{\mu_2\Gamma_2}{2}\big)\langle b_2\rangle-\big(iJ_3e^{i\theta_3}\notag
	\\&+\frac{\mu_3\Gamma_3}{2}\big)\langle b_2\rangle-i\varepsilon\notag
	\\
	\frac{d\langle b_1\rangle}{dt}=&-\frac{\Lambda_1}{2}\langle b_1\rangle-\big(iJ_1e^{-i\theta_1}+\frac{\mu_1^*\Gamma_1}{2}\big)\langle a\rangle\notag
	\\
	\frac{d\langle b_2\rangle}{dt}=&-\frac{\Lambda_2}{2}\langle b_2\rangle-\big(iJ_2e^{-i\theta_2}+\frac{\mu_2^*\Gamma_2}{2}\big)\langle a\rangle   \notag
	\\
	\frac{d\langle b_3\rangle}{dt}=&-\frac{\Lambda_3}{2}\langle b_3\rangle-\big(iJ_3e^{-i\theta_3}+\frac{\mu_3^*\Gamma_3}{2}\big)\langle a\rangle  ,
\end{align}
where $\mu_1=p_a^*p_{b_1}$, $\mu_2=p_a^*p_{b_2}$, $\mu_3=p_a^*p_{b_3}$ and $\Lambda_0=\kappa_{a}+(\Gamma_{1}+\Gamma_{2}+\Gamma_{3})|p_a|^2$, $\Lambda_1=\kappa_{b_1}+\Gamma_{1}|p_{b_1}|^2$, $\Lambda_2=\kappa_{b_2}+\Gamma_{2}|p_{b_2}|^2$, and $\Lambda_3=\kappa_{b_3}+\Gamma_{3}|p_{b_3}|^2$. Nonreciprocal coupling can be realized by setting $\theta_1=\theta_2=\theta_3=\pm\pi/2$, $p_a=p_{b_1}=p_{b_2}=p_{b_3}=\pm1$, and $J_i=\Gamma_i/2$.

The corresponding stored energies are
\begin{align}
	E_a(t)=&\frac{4\omega\varepsilon^2(1-e^{-\frac{1}{2}t\Lambda_0})^2}{\Lambda_0^2},\notag\\  E_{b_1}(t)=&\frac{16\omega\varepsilon^2|\mu_1|^2\Gamma_1^2}{\Lambda_0^2\Lambda_1^2(\Lambda_0-\Lambda_1)^2}\Big[(\Lambda_0-\Lambda_1)-(\Lambda_0 e^{-\frac{1}{2}\Lambda_1 t}\notag\\
    &-\Lambda_1 e^{-\frac{1}{2}\Lambda_0 t})\Big]^2,\notag\\  E_{b_2}(t)=&\frac{16\omega\varepsilon^2|\mu_2|^2\Gamma_2^2}{\Lambda_0^2\Lambda_2^2(\Lambda_0-\Lambda_2)^2}\Big[(\Lambda_0-\Lambda_2)-(\Lambda_0 e^{-\frac{1}{2}\Lambda_2 t}\notag\\
    &-\Lambda_2 e^{-\frac{1}{2}\Lambda_0 t})\Big]^2,\notag\\  E_{b_3}(t)=&\frac{16\omega\varepsilon^2|\mu_3|^2\Gamma_3^2}{\Lambda_0^2\Lambda_3^2(\Lambda_0-\Lambda_3)^2}\Big[(\Lambda_0-\Lambda_3)-(\Lambda_0 e^{-\frac{1}{2}\Lambda_3 t}\notag\\
    &-\Lambda_3 e^{-\frac{1}{2}\Lambda_0 t})\Big]^2.
\end{align}
Similar to the cascaded process, we can also summarize the recursive structure and give the
general energy distribution in Eq.~(\ref{eq7}).

The optimal effective coupling can be obtained by maximizing the steady-state stored energy with respect to the coupling strength $J$, i.e.,

\begin{align}
	\frac{\partial E_c^{\rm ss}(N)}{\partial J}=&-\frac{2^{3+2N}\omega\Gamma^{-1+2N}\varepsilon^2}{(\Gamma+\kappa)^5}(2\Gamma+\kappa)^{1-2N}(2\Gamma^2\notag\\
    &-N\Gamma\kappa-N\kappa^2)=0,
\end{align}	
and
\begin{align}
	\frac{\partial E_p^{\rm ss}}{\partial J}=\frac{32\omega\Gamma\varepsilon^2(-N\Gamma^2+\kappa^2)}{(\Gamma+\kappa)^3(N\Gamma+\kappa)^3}=0.
\end{align}	
which leads to the analytical expressions in the main text.

These results explicitly reveal how directional transport and cooperative dissipation determine the scaling of the stored energy with the battery number.
\section{Analytical derivation of the odd-even effect in the reciprocal cascaded configuration}\label{AB}

In this appendix, we analyze the origin of the energy difference between odd and even battery chains under reciprocal coupling. We consider a charger coupled to a chain of $N$ quantum batteries with nearest–neighbor reciprocal coupling, described by the Hamiltonian
\begin{equation}
H = \varepsilon (a + a^\dagger) 
+Ja^{\dagger}b_1+\sum_{i=2}^{N}Jb_{i-1}^{\dagger}b_{i}+\rm{H.c.}.\label{b1}
\end{equation}
Because the reciprocal configuration has an open-chain geometry, the hopping phases can be gauged away by local phase transformations, we therefore set $\theta_i=0$ for simplicity.

Including uniform dissipation $\kappa$, the amplitudes of the steady-state expectation values of the operators satisfy the set of linear equations
\begin{equation}
\begin{aligned}
\varepsilon &= \frac{i\kappa}{2}\alpha - J\beta_1, \\
0 &= \frac{i\kappa}{2}\beta_1 - J\,(\alpha + \beta_2), \\
0 &= \frac{i\kappa}{2}\beta_2 - J\,(\beta_1 + \beta_3), \\
&\;\;\;\vdots \\
0 &= \frac{i\kappa}{2}\, \beta_N - J\beta_{N-1},
\end{aligned}
\end{equation}
with $\alpha=\langle a\rangle_{\rm ss}$, $\beta_i=\langle b_i\rangle_{\rm ss}$.
Defining the vector $\mathbf{a} = (\alpha, \beta_1, \beta_2, \cdots, \beta_N)^T$, the steady state equation can be written as $(-H_0 + i\kappa/2)\, \mathbf{a} = \varepsilon \, \mathbf{e}_0$, where 
\begin{equation}
H_0 = 
\begin{pmatrix}
0 & J & 0 & \cdots & 0 \\
J & 0 & J & \cdots & 0 \\
0 & J & 0 & \cdots & 0 \\
\vdots & \vdots & \vdots & \ddots & J \\
0 & 0 & 0 & J & 0
\end{pmatrix}, \quad 
\mathbf{e}_0 = 
\begin{pmatrix}
1 \\ 0 \\ 0 \\ \vdots \\ 0
\end{pmatrix}.
\end{equation}
Here, $H_0$ is the tight-binding matrix associated with the Hamiltonian in Eq.~(\ref{b1}).

Defining the Green function as $G=(-H_0+i\kappa/2)^{-1}$, the steady-state amplitude of the $n$-th mode can be expressed as~\cite{economou2006green}
\begin{equation}
\beta_n = \varepsilon\, G_{n1}.
\end{equation}
Consequently, the energy stored in the $n$-th battery is determined by the corresponding Green-function matrix element $G_{n0}$.

We can expand the Green function in the eigenbasis of the Hamiltonian by introducing the completeness relation
\begin{equation}
G = \sum_k
\frac{1}{-E_k+i\kappa/2}|\psi_k\rangle\langle\psi_k|,
\end{equation}
with $E_k$ and $\psi_k$ are the eigenvalues and eigenvectors of $H_0$, $\sum_k |\psi_k\rangle \langle \psi_k| = I$. The corresponding matrix elements read
\begin{equation}
G_{ij} =
\sum_k
\frac{\psi_k(i)\psi_k(j)}
{-E_k + i\kappa/2},
\end{equation}
where $\psi_k(i)=\langle i|\psi_k\rangle$ and $\psi_k(j)=\langle \psi_k|j\rangle$.

Finally, the steady-state amplitude of the $n$-th battery is given explicitly by
\begin{equation}
\beta_n =
\varepsilon
\sum_k
\frac{\psi_k(n)\psi_k(0)}
{-E_k + i\kappa/2}.
\end{equation}

For an open chain of length $L=N+1$, the eigenmodes of the tight-binding Hamiltonian are given by
\begin{equation}
\psi_k(n)=\sqrt{\frac{2}{L+1}}
\sin\left(\frac{\pi k (n+1)}{L+1}\right),
\end{equation}
with corresponding eigenvalues
\begin{equation}
E_k = 2J\cos\left(\frac{\pi k}{L+1}\right),
\end{equation}
where $k=1,\dots,L$.
The eigenvector components at the first and last sites read
\begin{equation}
\psi_k(0)=
\sqrt{\frac{2}{L+1}}
\sin\left(\frac{\pi k}{L+1}\right),
\end{equation}
\begin{equation}
\psi_k(N)=
\sqrt{\frac{2}{L+1}}
\sin\left(\frac{\pi k L}{L+1}\right).
\end{equation}

Thus, using the Green-function expansion, the steady-state amplitude of the last battery is
\begin{equation}
\beta_N =
\frac{2\varepsilon}{L+1}
\sum_k
\frac{
\sin\left(\frac{\pi k}{L+1}\right)
\sin\left(\frac{\pi k L}{L+1}\right)
}{
-E_k+i\kappa/2
}.
\end{equation}
Applying the trigonometric identity
\begin{equation}
\sin\left(\frac{\pi k L}{L+1}\right)
=
(-1)^{k+1}
\sin\left(\frac{\pi k}{L+1}\right),
\end{equation}
we obtain
\begin{equation}
\beta_N =
\frac{2\varepsilon}{L+1}
\sum_k
\frac{
(-1)^{k+1}
\sin^2\left(\frac{\pi k}{L+1}\right)
}{
-E_k+i\kappa/2
}.
\end{equation}
The factor $(-1)^{k+1}$ introduces alternating signs in the contributions from different eigenmodes.

For chains with an even number of batteries, one eigenmode appears close to zero energy, i.e., $E_k = 0$,
which strongly enhances the corresponding Green function term
\begin{equation}
\frac{1}{-E_k+i\kappa/2}.
\end{equation}
This mode exhibits a large amplitude at the terminal sites, resulting in constructive energy transfer along the chain.

In contrast, for an odd number of batteries, the spectrum is symmetric around zero energy, and no eigenmode exists exactly at $E=0$. The contributions from different modes partially cancel due to the alternating sign $(-1)^{k+1}$, leading to destructive interference in the Green-function summation. Consequently, the steady-state amplitude at the last battery is smaller than in the even number case, i.e.,
\begin{equation}
|\beta_N|_{\text{even}} > |\beta_N|_{\text{odd}},
\end{equation}
which directly implies a larger stored energy.

\section{Analytical derivation of the steady-state energy in the reciprocal parallel configuration}\label{ACcc}

For the parallel configuration, the Hamiltonian is\begin{equation}
H = \varepsilon (a + a^\dagger) 
+\sum_{i=1}^{N}Ja^{\dagger}b_{i}+\rm{H.c.},\label{c1}
\end{equation}
where the coupling phase has been set to $\theta_i=0$ for simplicity.

Including local dissipation, the steady-state equation of the system operators is
\begin{equation}
\begin{aligned}
\varepsilon &= \frac{i\kappa}{2}\alpha - J(\beta_1+\beta_2+\dots+\beta_N), \\
0 &= \frac{i\kappa}{2}\beta_1 - J\,\alpha, \\
0 &= \frac{i\kappa}{2}\beta_2 - J\,\alpha, \\
&\;\;\;\vdots \\
0 &= \frac{i\kappa}{2}\, \beta_N - J\alpha,
\end{aligned}
\end{equation}
which can also be written as $(-H_0 + i\kappa/2)\, \mathbf{a} = \varepsilon \, \mathbf{e}_0$, with 
\begin{equation}
H_0 = 
\begin{pmatrix}
0 & J & J & \cdots & J \\
J & 0 & 0 & \cdots & 0 \\
J & 0 & 0 & \cdots & 0 \\
\vdots & \vdots & \vdots & \ddots & 0 \\
J & 0 & 0 & 0 & 0
\end{pmatrix},
\end{equation}
$\mathbf{a}=(\alpha,\beta_1,\cdots,\beta_N)^T$, and $\mathbf{e}_0=(1,0,\cdots,0)^T$.

Similarly, the Green function can be written as $G=(-H_0+i\kappa/2)^{-1}$, and the steady-state amplitude can be expressed as $\beta_n = \varepsilon\, G_{n0}$.

Unlike the cascaded configuration, the spectrum of $H_0$ has only two nonzero-energy eigenstates with eigenvalues \begin{equation}
E_{1,2}=\pm\sqrt{N}J,
\end{equation}
whose corresponding eigenstates are 
\begin{align}
|\psi_1\rangle&=\frac{1}{\sqrt{2N}}(\sqrt{N},1,1,\cdots,1)^T,\notag\\
|\psi_2\rangle&=\frac{1}{\sqrt{2N}}(-\sqrt{N},1,1,\cdots,1)^T.
\end{align}
The remaining $N-1$ eigenstates form a degenerate zero-energy dark-state manifold, and the corresponding eigenstates are
\begin{align}
|\psi_k\rangle&=
\frac{1}{\sqrt{2}}
(0,-1,0,\ldots,0,\underbrace{1}_{k-\text{th element}},0,\ldots,0)^T,\notag\\
&\qquad
k=3,\cdots,N+1.
\end{align}
Expanding the Green function in the eigenbasis yields
\begin{align}
G &= \sum_k
\frac{1}{-E_k+i\kappa/2}|\psi_k\rangle\langle\psi_k|\notag\\&=\sum_{k=1,2}
\frac{1}{-E_k+i\kappa/2}|\psi_k\rangle\langle\psi_k|+\sum_{k=3}^{N+1}
\frac{1}{i\kappa/2}|\psi_k\rangle\langle\psi_k|.
\end{align}

Since every dark state is orthogonal to the driving vector,
\begin{equation}
\langle\psi_k|\mathbf e_0=0,\qquad k=3,\cdots,N+1,
\end{equation}
their contributions vanish identically, and the steady-state response is completely determined by the two bright eigenmodes,
\begin{align}
\mathbf a
&=
\sum_{k=1,2}
\frac{\varepsilon}{-E_k+i\kappa/2}|\psi_k\rangle\langle\psi_k|\mathbf{e}_0.
\end{align}

After substituting $E_1$, $E_2$, and the corresponding eigenstates, we can obtain
\begin{equation}
\mathbf{a} = \frac{\varepsilon}{2}
\begin{pmatrix}
\frac{1}{-\sqrt{N}J+i\kappa/2}+\frac{1}{\sqrt{N}J+i\kappa/2} \\ \frac{1}{\sqrt{N}}[\frac{1}{-\sqrt{N}J+i\kappa/2}-\frac{1}{\sqrt{N}J+i\kappa/2}] \\ \vdots \\ \frac{1}{\sqrt{N}}[\frac{1}{-\sqrt{N}J+i\kappa/2}-\frac{1}{\sqrt{N}J+i\kappa/2}]
\end{pmatrix}.
\end{equation}
Thus, the steady-state energy storage of a single battery is
\begin{align}
    E_N^{\rm {ss}}=\omega|\beta_n|^2=\frac{16J^2\omega\varepsilon^2}{(4NJ^2+\kappa^2)^2}.
\end{align}

Unlike the cascaded configuration, the zero-energy dark modes in the parallel configuration carry zero spectral weight because they are orthogonal to the driving vector. Therefore, the energy transport is governed entirely by the two bright eigenmodes, independent of the parity of the battery number, and no odd-even effect appears. In the strong-coupling regime $J\gg\kappa$, the stored energy decreases asymptotically as $E_N^{\rm ss}\propto J^{-2}$.
\bibliography{REV}	
\end{document}